\documentclass[]{aipproc}
\layoutstyle{6x9}
\usepackage[english]{babel}
\usepackage{graphicx}
\newcommand{\eq}[1]{eq. (\ref{#1})}
\newcommand{\beq}{\begin{equation}}
\newcommand{\eeq}{\end{equation}}
\newcommand{\beqa}{\begin{eqnarray}}
\newcommand{\eeqa}{\end{eqnarray}}
\renewcommand{\(}{\left(}
\renewcommand{\)}{\right)}
\renewcommand{\d}{\mbox{d}}
\newcommand{\e}{\epsilon}
\newcommand{\g}{{\gamma}}
\newcommand{\m}{{\mathbf m}}
\newcommand{\x}{{\mathbf x}}
\newcommand{\y}{{\mathbf y}}
\newcommand{\z}{{\mathbf z}}
\newcommand{\1}{{\mathbf 1}}
\newcommand{\T}{{\mathbf T}}
\newcommand{\M}{{\mathbf M}}
\newcommand{\B}{{\mathbf B}}
\renewcommand{\t}[1]{#1^{\top}}
\newcommand{\inv}[1]{#1^{-1}}
\newcommand{\ave}[1]{\left\langle #1 \right\rangle}
\newcommand{\set}[1]{\left\{ #1 \right\}}

\begin{document}

\title{Random Feynman Graphs}

\author{B. S\"oderberg}{
  address={Complex Systems Division, Dept. of Theoretical Physics, Lund University, Sweden}
}

\date{\today}

\begin{abstract}
We investigate a class of random graph ensembles based on the Feynman
graphs of multidimensional integrals, representing
statistical-mechanical partition functions. We show that the resulting
ensembles of random graphs strongly resemble those defined in random
graphs with hidden color (CDRG), generalizing the known relation of
the Feynman graphs of simple one-dimensional integrals to random
graphs with a given degree distribution.
\end{abstract}

\pacs{02.50.-r, 64.60.-i, 89.75.Fb}

\keywords{graph theory; random graph; phase transition;
  critical phenomena; percolation; Feynman graph}

\maketitle

\section{Introduction}

For a long time, the concept of random graphs was synomymous with the
classic {\bf ER}-model, which has been extensively studied ever since
it was introduced in the sixties \cite{ErRe60,Jans00,Boll01}. In the
last decades, however, driven by the steadily increasing availability
of data on large real-world networks (e.g. various social, biological,
and information-technological networks), a large variety of more
general models have been considered; see e.g. \cite{DoMe03} for a
fairly up-to-date review.

One of the more commonly studied classes of models is {\em random
graphs with a given degree distribution}, or {\em the configuration
model} \cite{BenCan, MoRe95, Newm01}, here to be referred to as {\bf
DRG} (for ``degree-driven random graphs''). There, one considers a
maximally random graph with $N$ vertices, restricted such that its
degree sequence, $\{m_i, i=1\ldots N\}$, is consistent with a
specified degree distribution, $\{p_m\}$. Once the degrees of
individual vertices are fixed, edges are formed by means of a random
pairing of the entire set of stubs (half-edges), assuming an even
total stub-count. This leads in general to a multigraph, with multiple
edges and/or self-connections. The resulting graph ensembles are known
to lack non-trivial edge correlations.

Recently, extensions of these models based on adding so called {\em
hidden color}, have been studied. In a colored extension of the
ER-model, Inhomogeneous Random Graphs (IRG), vertices were assumed to
carry color variables, allowed to affect edge formation, which was
shown to result in a class of models allowing for non-trivial edge
correlations \cite{sod02}, while degree distributions were limited to
mixtures of Poissonians.

Similarly, colored extensions of the DRG class of models have been
considered, where not vertices but half-edges ({\em stubs}) were
considered to carry a color variable, represented e.g. by an integer
variable $c\in[1\ldots K]$, allowed to affect edge probabilities. The
degree of a vertex amounts to its stub count; since stubs are colored
it is natural to consider the concept of a {\em colored degree}, in
the form of a vector $\m = \{m_c,\, c=1\ldots K\}$, where $K$ is the
number of possible colors, and $m_c$ is the number of stubs of a
definite color $c$.  The conventional degree $m$ then is obtained
simply by summing up the components of its colored degree,
$m=\sum_{c=1}^K m_c$.
Thus, in the {\bf CDRG} (for {\em Colored DRG}) class of models
\cite{CDRG,CDRG2,CDRG3}, random graphs with a given {\em colored
degree distribution} are considered, where a sequence of colored
degrees is chosen consistently with a specified distribution. The
color is allowed to affect the edge distribution by introducing a
color-dependent bias in the random pairing of stubs. The color can
either be considered as an observable characteristic, or as unvisible
once it has done its job (hidden color).

In this article, we will study models of random graphs based on the
{\em Feynman graphs} associated with a class of simple statistical
mechanics models, where they represent the deviations of certain
integrals from a Gaussian approximation \cite{ItZu80}. Relations
between DRG models and the Feynman graphs of certain simple univariate
integrals are known, and have been explored e.g. in
refs. \cite{BuCoKr01,DoMeSa02}.

We will show that such relations are not unique for univariate
integrals; indeed, {\em any} integral of whatever dimension,
representing the partition function of a statistical-mechanical system
from a large class of models, will define an ensemble of random graphs
related to an associated CDRG model, and this is the main result of
this article.

One thing to be gained from such relations is an alternative, and
perhaps more natural, way to define the associated DRG or CDRG random
graph ensembles, in terms of random Feynman Graphs. 

In addition, as always when relations are found between different
areas of science, one may expect conceptual gains, where phenomena in
one area can be simply understood in terms of well-known phenomena in
the other. Thus, the relations raise questions, e.g., as to how
critical parameter values in the random graph models are mirrored in
terms of phase transitions in the associated statistical physics
models.

\section{Random Feynman Graphs - univariate case}

In field theory and statistical physics, {\em Feynman graphs} provide
a standard method for organizing the perturbative expansion of an
integral $Z$ representing some field theoretical or statistical
physics model, around an approximate value as given by a Gaussian
approximation to the integrand (see e.g. \cite{ItZu80}).

Sometimes this approximation is based on a saddle point, an extremal
point of the integrand, with the Gaussian chosen to approximate the
integrand well in the neighborhood of the saddle point. However, any
decomposition of the integrand into the product of a Gaussian factor
and a non-trivial remainder will do.

In the spirit of refs. \cite{BuCoKr01,DoMeSa02}, we will consider the
possibility of defining ensembles of random graphs -- {\em random
Feynman graphs} -- based on the Feynman graphs for such a system, with
the statistical weight of each distinct graph taken to be proportional
to the corresponding term in the perturbative expansion of $Z$. Much
of the contents of this section concerns known results
\cite{BuCoKr01,DoMeSa02}, included to make the article reasonably
self-contained.

\subsection{Feynman Graphs for a Simple Integral}

In the simplest case, the integral would be over a single real
variable, $x$, say, with an integrand in the form of the exponential
of an action $S(x)$, that can be written as the sum of a simple
quadratic term $S_0(x)$ that defines the {\em unperturbed action}, and
a non-trivial {\em perturbation} $V(x)$, typically a more complicated
function. Thus, consider the (formal) integral
\beq
	Z = \int e^{S(x)} \d x
	= \int e^{S_0(x) + V(x)} \d x
	= \int e^{-bx^2/2} e^{V(x)} \d x
\eeq
where we restrict ourselves to $b>0$, and to a perturbation $V$ with
only non-negative Taylor coefficients,
\beq
	V(x) = \sum_n v_n \frac{x^n}{n!},
	\;\mbox{with}\;\;
	v_n\ge 0,
	\;\mbox{for}\;n=0,1\dots
\eeq
The Feynman graphs for such an integral represent different terms in
the perturbative expansion of $Z$ around the Gaussian approximation,
$Z_0 = \int e^{S_0(x)} \d x = \int e^{-bx^2/2} \d x$.

To see this, first note that considering $V(x)$ as a perturbation
means that it is to be considered as small in some sense, and so it is
natural to write $Z$ as the sum of a sequence of contributions,
obtained by expanding the non-trivial factor $e^{V(x)} = \sum_n
V(x)^n/n!$, yielding
\beq
	Z = \sum_N \frac{Z_N}{N!},\;\;\mbox{with}\;\; Z_N = \int e^{-bx^2/2} V(x)^N \d x.
\eeq
Now, each of the $N$ factors of $V(x)$ is to be associated with a
distinct {\em vertex}, labelled $i=1,\dots,N$. Next, by
Taylor-expanding each factor of $V(x)$, $Z_N$ decomposes into a sum of
contributions,
\beqa
	Z_N &=& \sum_{\set{m_i}} Z_{\set{m_i}} , \; \mbox{with}
\\
	Z_{\set{m_i}} &=& \prod_i \frac{v_{m_i}}{m_i!} \int e^{-bx^2/2} \prod_i x^{m_i} \d x
\eeqa
where the sequence $\set{m_i,\,i=0,\dots,N}$ represents the chosen terms
$\set{v_{m_i}x^{m_i}/m_i!}$ in the respective Taylor expansions.

Now each factor $x^{m_i}$ in the integrand represents a {\em vertex}
of {\em degree} $m_i$, with each single factor of $x$ representing a
{\em stub}. Each term is now easy to evaluate. Denoting by $M$ the
total degree $M=\sum_i m_i$, we have $\int e^{-bx^2/2} x^M \d x =
\e_M b^{-M/2} (M-1)!!$, where $\e_M$ is a 0/1-variable
enforcing the restriction to even $M$;
\beq
	\e_M \equiv \frac{1 + (-1)^M}{2}.
\eeq
The resulting factor of $(M-1)!!$ has an obvious combinatorial
interpretion as the {\em total number of distinct pairings} of the $M$
factors of $x$.  In each pairing, each paired couple of $x$-factors
can be associated with an {\em edge} between the corresponding
vertices $i,j$.

By considering the $m_i$ distinct factors of $x$ associated with the
stubs of a vertex as indistinguishable, pairings can be naturally
grouped into equivalence classes. Each such class of equivalent
pairings can be associated with a distinct graph, a {\em Feynman
graph} $\g$, representing a specific contribution to $Z$, or
rather to $\hat{Z} = Z/Z_0$.
\beq
	\hat{Z} = \sum_{\g} Z_{\g}.
\eeq
The contribution $Z_{\g}$ to the total $\hat{Z}$ from a arbitrary
Feynman graph $\g$ defines the {\em value} of the graph, which is
given by the following {\em Feynman rules}.
\begin{enumerate}
\item Each vertex $i$ is associated with a factor of $v_{m_i}$, where
$m_i$ is its degree;
\item Each edge is associated with a factor of $b^{-1}$;
\item The vertex and edge factors are multiplied together, and the
 result divided by the proper edge symmetry factor $S_{\g}$ (and
 of course by $N!$).
\end{enumerate}
The edge symmetry factor $S_{\g}$ is the usual one, accounting for
teh order of the edge symmetry group, with a factor of $n!$ for each
$m$-fold edge between distinct vertices, and a factor of $2^nn!$ for
each vertex with $n$ self-connections (where the $2^n$ comes from the
symmetry under flipping edges involved in self-connections).

The above rules apply to {\em labelled graphs}, where vertices are
considered distinguishable, while edges are indistinguishable, as are
their directions.  In many cases one considers instead {\em unlabelled
graphs}; the only difference is that then, also vertices are
considered indistinguishable. As a result, the $1/N!$ factor will be
replaced by the correct vertex symmetry factor of the particlar graph
considered, which together with the edge symmetry factor yields a {\em
total symmetry} factor $\hat{S}_{\g}$.

We will, however, focus on labelled graphs, that have distinguishable
vertices.

\subsection{Saddlepoints and Degree Distributions}

Following refs. \cite{BuCoKr01,DoMeSa02}, we now wish to interpret the
Feynman graphs associated with a particular integral as {\em random
graphs}, with a statistical weight proportional to the corresponding
contribution to $\hat{Z}$, as given by the Feynman rules above.

Then it is interesting to investigate what kind of random graph
ensembles this leads to. For a fixed $N$, it is obvious that the
probability for a certain {\em degree sequence}, $\set{m_i}$, is given
by
\beq
	P\set{m_i} = \frac{Z_{\set{m_i}}}{Z_N}
\eeq
If we wish to focus on the degree of a single vertex, all we have to
do is to sum over the $N-1$ other degrees, to obtain the single vertex
{\em degree distribution}
\beq
	p_m = \frac{v_m}{m!}
	\frac{
	  \int e^{-bx^2/2} x^m V(x)^{N-1} \d x
	}{
	  \int e^{-bx^2/2} V(x)^N \d x
	}
\eeq
having the {\em generating function}
\beq
	H(z) \equiv \sum_m p_m z^m =
	\frac{
	  \int e^{-bx^2/2} V(xz) V(x)^{N-1} \d x
	}{
	  \int e^{-bx^2/2} V(x)^N \d x
	}
\eeq

Now, assume for large $N$ that $Z_N$ can be evaluated in a saddlepoint
approximation, obtained from the neighborhood of a real, positive
saddle point $x=x_0$, satisfying the saddle point condition
corresponding to an extremal value of the ``effective action'',
$S_N(x) = -bx^2/2 + N\log V(x)$,
\beq
\label{sp_cond}
	S_N'(x_0) \equiv b x_0 - N V'(x_0)/V(x_0) = 0
\eeq
To leading order, the saddle point approximation yields
\beq
	Z_N \propto e^{-x_0^2/2} V_0^N,
\eeq
with $V_0=V(x_0)$, yielding
\beq
	p_m \approx \frac{v_mx_0^m}{m!V_0}
	\;\;\Leftrightarrow\;\;
	H(z) \approx \frac{V(x_0z)}{V_0}.
\eeq
Thus, the perturbation $V(x)$ has an obvious relation to the
generating function for the single-vertex degree distribution of
Feynman graphs; this has been pointed out and explored to some extent,
e.g. in ref. \cite{BuCoKr01}, where the perturbation was set to $H$
from the start. Note, however, that {\em any} perturbation $V$ with
non-negative Taylor coefficients will do; by rescaling the integration
variable, $V$ can be recast into the form of $H$.

We can also consider the joint distribution of the degrees of a small
number of vertices. Thus, for the pair distribution, we obtain the
nicely factorizable bivariate generating function,
\beq
	H_2(z_1,z_2) \approx \frac{V(x_0z_1)V(x_0z_2)}{V(x_0)^2} \approx H(z_1) H(z_2)
\eeq
This factorization generalizes to larger sets, and indicates the
approximate {\em independence} of the degree distributions of distinct
vertices.

To clear the view, let us change the integration variable to one
rescaled by $x_0$: with $y = x/x_0$ and $V_0 = V(x_0)$, the saddle
point condition (\ref{sp_cond}) yields $b = N V'(x_0) / x_0 V_0$, and
we obtain, apart from an uninteresting constant,
\beq
\label{ZN}
	Z_N = \int \exp\(- \frac{N \bar{m}}{2} y^2 \) H(y)^N \d y
\eeq
with an integrand designed to have an extremum at $y=1$; $\bar{m}$ is
defined as the expected average degree, as given by $\bar{m} = H'(1)$,
while $H(1)\equiv\sum_m p_m = 1$ of course is assumed. Note that the
integrand has the form of a fixed factor taken to the $N$th power,
indicating that for large $N$, the saddle point approximation should
be OK (provided the correct saddle point is used).

In what follows, we will assume we have performed such a rescaling to
a natural variable $y$, and that $Z_N$ is defined by \eq{ZN}, with
$H(z)=\sum_m p_m z^m$ taken to generate a {\em desired degree
distribution} $\set{p_m}$.

\subsection{Similarities and Differences to DRG}

Based on the expression \eq{ZN} for $Z_N$, let us compare the
resulting ensemble of {\em Random Feynman Graphs} (RFG) and compare it
to the corresponding DRG model with a degree distribution $\set{p_m}$
given by the one generated by $H(x)$.

Let us first note that Feynman graphs are based on Gaussian integrals,
where the value $(M-1)!!$ of the integral $\int x^M e^{-bx^2/2}\d x$
for an even $M$ obviously stems from the number of distinct pairings
of the $M$ factors of $x$ in $x^M$. Thus, for a given degree sequence
$\set{m_i}$ yielding an even total stub count $M=\sum_i m_i$, the
distribution over compatible Feymnan graphs is obviously given by
random stub pairing, just as in DRG. Thus, the {\em restrictions to a
fixed degree sequence} of a DRG model and the associated RFG model are
obviously identical, and so it is enough to compare the distributions
over degree sequences, $P\set{m_i}$. In DRG, this factorizes as
$\prod_i p_{m_i}$, with the small modification that $M$ must be even.

For RFG, we obtain, with $M\equiv\sum_i m_i$,
\beq
	P\set{m_i}
	= \prod_i p_{m_i} \frac{
	  \int e^{-\frac{N\bar{m}}{2} x^2} x^M \d x
	}{
	  \int e^{-\frac{N\bar{m}}{2} x^2} H(x)^N \d x
	}
\approx \prod_i p_{m_i} (M-1)!! (N\bar{m})^{-M/2})\frac{
	  \int e^{-\frac{N\bar{m}}{2} x^2} x^M \d x
	}{
	  \int e^{-\frac{N\bar{m}}{2} x^2} H(x)^N \d x
	}
\eeq
which, apart from a factor depending only on $M$, is identical to the
DRG expression. This implies that the {\em restrictions to a fixed
total stub count $M$} of a DRG model and the associated RFG model are
{\em identical}, and so it is enough to compare the distributions over
total stub counts, $P_M$.

The $M$-distribution is simplest expressed in terms of its generating
function, $U(z) = \sum_M P_M z^M$. For DRG, it is essentially given by
$U(z) \sim H(z)^N$. The trivial requirement of an even $M$ yields the
slightly modified expression
\beq
	U^{\mbox{\tiny DRG}}(z) = \frac{H(z)^N + H(-z)^N}{1 + H(-1)^N}
\eeq
corresponding to the distribution
\beq
	P^{\mbox{\tiny DRG}}_M \propto
	\e_M \sum_{\set{m_i}} \delta_{M,\sum m_i} \prod_i p_{m_i}
\eeq
For RFG, on the other hand, the result is obviously given by
$P^{\mbox{\tiny RFG}}_M = \frac{Z_{NM}}{Z_N}$, yielding the generating
function
\beq
	U^{\mbox{\tiny RFG}}(z) =
	\frac{
	  \int e^{-\frac{N\bar{m}}{2}x^2} H(xz)^N \d x
	}{
	  \int e^{-\frac{N\bar{m}}{2}x^2} H(x)^N \d x
	}
\eeq
The resulting $M$-distribution can be written as
\beq
\label{comp}
	P^{\mbox{\tiny RFG}}_M
	\propto
	(M-1)!! \(N\bar{m}\)^{-M/2} P^{\mbox{\tiny DRG}}_M,
\eeq
and that is essentially the {\em only} difference between the models.

Thus, upon clamping the total stub count $M$ to an even number close
to the expected value $\bar{M} = N\bar{m}$, a pair of associated DRG
and RFG models are identical \cite{BuCoKr01,DoMeSa02}.  Note that the
relative factor in \eq{comp}, $\(N\bar{m}\)^{-M/2} P^{\mbox{\tiny
DRG}}_M$, is stationary for the expected value $M = N\bar{m}$, which
shows that the agreement for clamped $M$ is not very sensitive to the
precise clamped value, as long as it is not too far from the expected
value $\bar{M}$.

\subsection{Validity of the Saddle Point Approximation}

At the basis of the above discussion is the assumption that the saddle
point contributions really dominates $Z_N$ for large $N$, which is not
obviously always the case. A necessary condition is that $x_0=1$
defines a {\em local maximum} of the effective action $S_{\mbox{\tiny
eff}}(x) \propto -\bar{m}x^2/2 + \log H(x)$, yielding the rather weak
requirement
\beq
	S''_{\mbox{\tiny eff}}(x=1) \propto \ave{m(m-2)} - \ave{m}^2 < 0
	\;\;\Leftrightarrow\;\;
	\ave{m^2} - \ave{m}^2 < 2\ave{m}
\eeq
This does not appear to correspond to any obvious property of the
associated graph ensemble.

In case this condition is not fulfilled, the contributions from $x
\approx 1$ will fail to dominate $Z_N$ for large $N$. A possible
scenario then is that a competing saddle point $x_0\ne 1$ will take
over; assume it is on the positive real axis. Then we can always
define a {\em rescaled} integration variable $\tilde{x}=x/x_0$, such
that the proper saddle point is moved to $\tilde{x}=1$. The resulting
degree distribution $\set{\tilde{p}_m}$ will have a geometric factor
as compared to the expected one,
\beq
	\tilde{H}(\tilde{x}) = \frac{H(x_0 \tilde{x})}{H(x_0)}
	\;\;\Leftrightarrow\;\;
	\tilde{p}_m = \frac{x_0^m p_m}{H(x_0)}.
\eeq

Similarly, if we consider the full $Z$ instead of merely the
restriction $Z_N$ to a fixed order $N$, we have, in the rescaled
variables,
\beqa
	Z &\propto& \int e^{N_0 s(x)} \d x,\;\; \mbox{with}
\\
	s(x) &=& -\bar{m} \frac{x^2}{2} + H(x)
\eeqa
where the parameter $N_0$ is to be taken as a {\em desired value} of
$N$. Then $s'(1) = 0$, and we have a saddle point at $x=1$, just as
for $Z_N$ with $N=N_0$. This time, however, the requirement of a local
maximum, $s''(1) < 0$, is stronger; furthermore, it has an obvious
interpretation: it becomes the well-known condition of {\em
subcriticality}, i.e. the non-existance of a {\em giant component} in
the associated DRG model,
\beq
	-\bar{m} + H''(x) < 0
	\Leftrightarrow
	\ave{m(m-2)} < 0
\eeq
%

It is also interesting, for a fixed large $N_0$, to compare the full
$Z$ to the restrictions $Z_N$ for different $N$ in the neighborhood of
$N_0$, and to the double restrictions $Z_{NM}$ for different $M$ in
the neighborhood of $M_0 = N_0\bar{m}$. Of course, $Z$ will almost
always define a divergent integral, but we can assume it to be
formally dominated by the saddle point value, and compare that to the
contributions from different $N$ and $M$. We have
\beqa
	Z &=& \int e^{-N_0 \bar{m} x^2/2 + N_0 H(x)} \d x
\\
	Z_N &=& \frac{N_0^N}{N!} \int e^{-N_0 \bar{m} x^2/2} H(x)^N \d x
\\
	Z_{NM} &=& \frac{N_0^N}{N!} \oint \frac{\d z}{2\pi iz} z^{-M} H(z)^N
	\int e^{-N_0 \bar{m} x^2/2} x^M \d x
\\
\eeqa
where we assume $Z$ to have a proper saddle point at $x=1$,
i.e. $\ave{m(m-2)} < 0$. This implies the saddle point value
$e^{N_0(1-\bar{m}/2}$ for $Z$, as well as for $Z_N$ with $N=N_0$.  For
$Z_N$ with a slightly different $N$, the value will to lowest order
not change -- the prefactor is obviously staionary, and due to the
dominance of the saddle point $x=1$, a slightly different power of
$H(1)=1$ makes no difference.

Thus, $Z_N$ as a function of $N$ will have an extremum at $N=N_0$;
furthermore, the corresponding saddle point value coincides will that
of $Z$. In a similar way, $Z_{NM}$ as a function of $M$ for fixed
$N=N_0$, will have an extremum for $M=N\bar{m}$, and the saddle point
value coincides with that of $Z_N$. This indicates that under certain
circumstances, $Z$ gets its most important contributions from $N \sim
N_0$ and $M \sim N_0\bar{m}$, as it should.

\section{Random Feynman Graphs - multivariate case}

Now let us investigate instead the slightly more complicated case of a
multivariate integral, representing a statistical mechanical partition
function for a $K$-dimensional variable $\x = \set{x_a,\;a=1\dots
K}$. Thus, consider the $K$-dimensional integral
\beq
	Z
	= \int e^{S(\x)} \d^K\x
	= \int e^{ S_0(\x) + V(\x)} \d^K\x
	= \int e^{ -\frac{1}{2} \t{\x} \B \x} e^{V(\x)} \d^K\x
\eeq
with an action $S(\x)$ in the form of an unperturbed quadratic action
$S_0(\x)=-\sum_{ab} x_a B_{ab} x_b/2$ plus a perturbation $V(\x)$.  We
will restrict our attention to cases where (i) the inverse $\inv{\B}$
of the symmetric matrix $\B$ has no negative elements, and (ii) the
perturbation $V(\x)$ has only non-negative multivariate Taylor
coefficients, i.e.
\beqa
	&& V(\x) = \sum_{\set{m_a}} v_{\set{m_a}} \prod_a \frac{x_a^{m_a}}{m_a!} = \sum_{\m} v_{\m} \frac{\x^{\m}}{\m!},
\\
	&& \mbox{with} \;\;\; v_{\m} \ge 0
\eeqa
in terms of the multivariate power $\m = \set{m_a,a=1\dots K}$, and $\m!\equiv\prod_a m_a!$.

\subsection{Feynman Graphs for a Multivariate Integral}

The partition function $Z$ can be evaluated in a perturbative manner
around the Gaussian approximation, in complete analogy to the
univariate case, with the perturbative expansion organized in the
form of a sum of contributions associated with Feynman graphs, as
follows.

The first step is to expand $e^{V(\x)}$, yielding the decomposition
\beqa
	Z = \sum_N \frac{Z_N}{N!},\; \mbox{with}
\\
	Z_N = \int e^{-\frac{1}{2} \t{\x} \B \x} V(\x)^N \d^K\x
\eeqa
Then, each of the $N$ factors $V(\x)$ is associated with a distinct
vertex $i$, and Taylor-expanded to yield
\beqa
	Z_N &=& \sum_{\set{\m_i}} Z_{\set{\m_i}} ,\; \mbox{with}
\\
\label{Z_c_m}
	Z_{\set{\m_i}} &=& \prod_{i=1}^N \frac{v_{\m_i}}{\m_i!}
	\int e^{-\frac{1}{2} \t{\x} \B \x}
	\prod_i \x^{\m_i}
	\d^K\x
\eeqa
For each of the factors $\x^{\m_i}$, its multivariate power $\m_i$ is
interpreted as the {\em colored degree} of vertex $i$; thus, each of
its $m_{ia}$ factors $x_a$ is associated with a stub with color $a$,
belonging to vertex $i$, which has the (plain) degree $m_i=\sum_a
m_{ia}$.

It is obvious that the integral in \eq{Z_c_m} depends on $\set{\m_i}$
only via the total colored stub count $\M=\sum_i \m_i$. Again, as in
the univariate case, the value can be seen as arising from all
possible ways to pair the $M=\sum_a M_a$ (which again must be even)
distinct $\x$-factors.

The pairings can be organized in equivalence classes, corresponding to
distinct {\em Feynman graphs}, as usual characterized by a unique
adjacency matrix $\set{n_{ij}}$.  Each element $n_{ij}$ counts the
number of stubs of vertex $i$ that are paired with a stub from vertex
$j$. The value of $\hat{Z} = Z/Z_0$, with $Z_0=\int e^{-\frac{1}{2}
\t{\x} \B \x}\d x$, becomes the sum over contributions from the
individual Feynman graphs $\g$,
\beq
	\hat{Z} = \sum_{\g} Z_{\g}
\eeq
These counts can be further decomposed into the elements of a {\em
colored adjacency matrix} $\set{n_{iajb}}$, with elements $n_{iajb}$
counting the number of $a$-colored stubs of vertex $i$ that are paired
with a $b$-colored stub of vertex $j$, such that the sum rules
$m_{ia}=\sum_{jb} n_{iajb}$ and $n_{ij}=\sum_{ab} n_{iajb}$ hold.

Thus, the contribution to $\hat{Z}$ from each graph $\g$ -- its
{\em value} $Z_{\g}$ -- is due to a sum over pairings consistent
with that graph, and is given by the following {\em Feynman rules} for
the contribution.
\begin{enumerate}
\item Each stub in $\g$ is assigned an independent color variable
$a$.
\item Each vertex $i$ is associated with a factor $v_{\m_i}$,
where $\m_i=\set{m_{ia}}$ is its colored degree;
\item Each edge with stub colors $(a,b)$ at its endpoints is
associated with a factor of $\inv{\B}_{ab}$;
\item The vertex and edge factors are multiplied together, the result
is summed over all stub colors, and the result divided by the proper
edge symmetry factor $S_{\g}$ (and of course by $N!$).
\end{enumerate}
These rules are obvious generalizations of the univariate versions.

Now we want to interpret these Feynman graphs as {\em random graphs}
in a multivariate version of {\em RFG}, with a statistical weight for
each distinct graph proportional to the corresponding graph value as
given by the Feynman rules.

\subsection{Saddlepoints and Colored Degree Distributions}

We wish to investigate the properties of the resulting graph
ensemble. To that end, let us investigate the degrees, or even better,
the {\em colored degrees}, of the resulting graphs. For a fixed $N$,
the distribution over colored degree sequences $\set{\m_i}$ is
obviously given by
\beq
	P\set{\m_i} = \frac{Z_{\set{\m_i}}}{Z_N}
\eeq
By summing over the colored degrees of all vertices except one, the
colored degree distribution of a single vertex results, and is given
by
\beq
	p_{\m} = \frac{v_{\m}}{\m!}
	\frac{
	\int e^{-\frac{1}{2} \t{\x} \B \x}
	\x^{\m} V(\x)^{N-1}
	\d^K\x
	}{
	\int e^{-\frac{1}{2} \t{\x} \B \x}
	V(\x)^N
	\d^K\x
	}
\eeq

As was the case for univariate integrals, we can hope to evaluate this
with {\em saddle point} methods in the thermodynamic limit of
$N\to\infty$. For a point $\x_0$ to be a saddle point of $Z_N$, it
must render the integrand of $Z_N$ extremal, corresponding to the
saddle point condition
\beq
	0 = -\B \x_0 + N \frac{\nabla V}{V}
	\;\;\Rightarrow\;\;
	\B \x_0 V(\x_0) = N \nabla V(\x_0)
\eeq
For the single-vertex colored degree distribution, the saddle point
approximation yields
\beq
	p_{\m} = \frac{v_{\m}\x_0^{\m}}{\m! V(\x_0)}
\eeq
with the generating function $H(\z) = \sum_{\m} p_{\m} \z^{\m}$
given by
\beq
	H(\z) = \frac{ V(\x_0\circ\z)}{V(\x_0)}
\eeq
where the ring ``$\circ$'' represents the componentwise multiplication
of two vectors.

Assuming the saddle point $\x_0$ to have entirely positive components,
$x_{0,a}>0$, we can change integration variables to rescaled variables
$\y$, given by $\x=\y\circ\x_0$. Accordingly, we can define
$V_0=V(\x_0)$, in terms of which $V(\x)=V(\y\circ\x_0)=V_0 H(\y)$. We
also choose to define a transformed matrix $\T$ from $N\inv{\T} =
\x_0\circ\B\circ\x_0$. yielding $T_{ab}\ge 0$.

For $Z_N$, this yields the expression
\beq
\label{ZNc}
	Z_N \propto
	\int e^{-\frac{N}{2} \t{\y} \inv{\T} \y} H(\y)^N \d^K\y
\eeq
with an integrand constructed to have a saddle point at
$\y=\1=\set{1,\dots,1}$, which implies the following constraint on the
matrix $\T$.
\beq
\label{T_con}
	\inv{\T} \1 = \frac{\nabla H(\1)}{H(\1)}
	\;\; \Rightarrow \;\;
	\T\bar{\m} = \1
\eeq
where we have introduced the {\em average colored degree},
$\bar{\m}=\nabla H|_{\y=\1}$, while normalization of the colored
degree distribution with necessity implies $H(\1) = 1$.

\subsection{Similarities and Differences to CDRG}

Now let us assume we are using the convenient coordinates $\x$, with
the saddle point asymptotically assumed to be at $\x=\1$, in terms of
which $Z_N$ is defined by \eq{ZNc}.

From the above discussion, it is should be obvious that the
multivariate RFG approach shows strong similarities to the
corresponding CDRG model \cite{CDRG2}, which is defined in terms of a
given graph size $N$, a color set $[1,\dots,K]$, a {\em colored degree
distribution} $\set{p_{\m}}$, and a {\em color preference matrix},
$\T=\set{T_{ab}}$, with non-negative elements $T_{ab}$, required to
fulfill the constraint \eq{T_con}, as follows. For each vertex, its
colored degree is drawn independently from the given
distribution. Then a {\em weighted random pairing} of the full set of
$M=\sum_a M_a$ colored stubs is performed, where each edge connecting
colors $a,b$ provides a weight factor $\propto T_{ab}$.

For the random Feynman graphs, the distribution over colored degree
sequences, for fixed $N$, becomes
\beq
\label{Pc_RFG}
	P^{\mbox{\tiny RFG}}_{\set{\m_i}} = \frac{Z_{\set{\m_i}}}{Z_N}
	=
	\prod_i p_{\m_i}
	\frac{
	  \int e^{-\frac{N}{2} \t{\x} \inv{\T} \x} \x^{\M} \d^K\x
	}{
	  \int e^{-\frac{N}{2} \t{\y} \inv{\T} \y} H(\y)^N \d^K\y
	}
\eeq
This can be compared to the corresponding CDRG result, which of
course has the factorized structure
\beq
\label{Pc_CDRG}
	P^{\mbox{\tiny CDRG}}_{\set{\m_i}}
	=
	\prod_i p_{\m_i}
\eeq
In analogy to the univariate case, we note that the associated RFG and
CDRG ensembles agree on (1) the distribution over colored graphs
conditional on a fixed colored degree sequence, and (2) the
distribution over colored degree sequences conditional upon a fixed
total colored stub count $\M$. The former follows from the equivalence
of the random pairing step involved in both ensembles, and the latter
from a direct comparison of eqs. (\ref{Pc_RFG},\ref{Pc_CDRG}).

Thus, the two ensembles only disagree on the distribution over the
total colored stub count $\M$; once it is clamped to a certain value,
the two yield identical distributions over graphs, even before summing
over colors.

In agreement with the univariate case, the condition for the saddle
point at $\x=\1$ to define a local maximum of the integrand for the
formal integral $Z = \int e^{N_0\(-\t{\x}\inv{\T}\x/2+H(\x)\)}\d^K\x$
is equivalent to the condition of {\em subcriticality} of the
associated CDRG model \cite{CDRG2}. Also, $Z$ is formally dominated by
graphs with $N=N_0$ and $\M=N_0\bar{\m}$, in the sense that the saddle
point values of $Z$ and the restrictions $Z_N$ and $Z_{N\M}$ all agree
to leading order, while $Z_N$ as a function of $N$ is stationary for
$N=N_0$, as is $Z_{N_0\M}$ as a function of $\M$ for $\M=N_0\ave{\m}$.

\section{Conclusions}

A large class of ensembles of Feynman graphs for integrals
representing statistical-mechanical partition functions have been
investigated as models for random graphs, with the statistical weight
of a graph taken to be proportional to its associated value as a
contribution to the integral.

For the case of univariate integrals, it was previously known
\cite{BuCoKr01} that the resulting random Feynman graph ensembles
closely resemble those of random graphs with a given degree
distribution, or DRG.
In this article, it was found that for the case of {\em multivariate}
integrals, the associated random Feynman graphs in a similar way
resemble those of CDRG, a stub-colored extension of DRG. These
findings are new, although such relations were conjectured in
ref. \cite{CDRG2})

For the restriction of the ensembles to a clamped value of the total
(colored) stub count, the agreement was in fact shown to be exact,
while the distribution over the total (colored) stub count was shown
to differ, which leads to differences also in the resulting degree
distributions. We speculate that this may be the effect of the random
Feynman graph ensemble defining a kind of {\em annealed approximation}
to the (C)DRG one, where the disorder as defined by the random
(colored) degree sequence is considered {\em quenched}, i.e. randomly
chosen from a given distribution, then clamped in the random pairing
step. In the random Feynman graphs, as a contrast, both degrees and
pairings fluctuate on an equal level.

The establishment of close relations between different models opens up
the stage for questions as to the relation between the critical
phenomena in the graph models (such as the appearance of a giant
component) and those in the associated statistical-mechanical
models. Some relations of this type was found, but more work is needed
to further illuminate this issue.

Finally, we note that a straightforward generalization to random
Feynman graph models for {\em directed} stub-colored graphs should be
possible by considering suitably defined multidimensional integrals
over a set of {\em complex} variables.

\section*{Acknowledgments}

It is a pleasure to thank the organizers of CNET2004 for the
invitation, and for a well organized conference in the charming city
of Aveiro. This work has in part been supported by the Swedish
Foundation for Strategic Research.

\bibliographystyle{aipprocl}
%
%

\end{document}